\documentclass[prl,aps,floats,twocolumn]{revtex4}

\usepackage{graphics}
\usepackage{amssymb}

\begin{document}
            
\title{Competition between Hidden Spin and Charge Orderings
in Stripe Phase}

\author{Chitoshi Yasuda$^{1,2}$, Akihiro Tanaka$^1$, and Xiao Hu$^1$}

\affiliation{$^1$ Computational Materials Science Center, National
Institute for Materials Science, Tsukuba 305-0047, Japan}
\affiliation{$^2$ Department of Physics and Mathematics, Aoyama Gakuin
University, Sagamihara 229-8558, Japan}

\date{\today}

\begin{abstract}
The correlation between charge and spin orderings in hole-doped
antiferromagnets is studied within an effective model of quantum strings
fluctuating in an antiferromagnetic background. In particular, we
perform the direct estimation of the charge and spin long-range-order
parameters by means of the quantum Monte Carlo simulation.
A hidden spin long-range order is found to be governed by a competition
 between the two trends caused by increasing hole mobility: the
 enhancement of the two-dimensional spin-spin correlation mediated by
 hole motions and the reformation of a strong stripe order. 
\end{abstract}

\pacs{74.25.Dw}

\maketitle
Since the observation of high-temperature superconductivity in a
hole-doped antiferromagnetic (AF) insulator, spin dynamics in hole-doped
low-dimensional Heisenberg antiferromagnets has turned into a major
subject of theoretical and experimental studies. The discovery of an
anomalous stripe phase within CuO$_2$ planes of the Cu-oxide
superconductor, La$_{2-x-y}$Nd$_y$Sr$_x$CuO$_4$, has stimulated
considerable interest in the studies of the correlation between the
charge and spin orderings in hole-doped
antiferromagnets~\cite{Tranquada1}. The experimental findings suggest
that, in the stripe phase, the dopant-holes are segregated into
one-dimensional (1D) domain walls which separate AF anti-phase spin
domains~\cite{Noda,Zhou}. The stripe phases have been experimentally
observed in a wide range of doped-hole concentrations for the lanthanum
cuprate~\cite{Cheong,Ino} and 
proposed in the two-dimensional (2D) Hubbard-like
~\cite{Machida,Poilblanc,Zaanen3,Schulz,Giamarchi,Miyazaki} and
$t$-$J$ models~\cite{White, Sakai}.
It is generally accepted that the
static stripe order, e.g., observed for La$_{2-x}$Sr$_x$CuO$_4$ with $x
\simeq 1/8$, is pinned by the lattice modulation in the low-temperature
tetragonal phase.

The stripe order with AF anti-phase spin domains has also
been observed
in Ni-oxide La$_{2-x}$Sr$_x$NiO$_4$~\cite{CChen, Tranquada3}, though
the material remains insulating except at very high Sr concentrations. 
There is by now a wide consensus that the formation of stripe is a  
generic property of hole-doped antiferromagnets~\cite{Tranquada1}. The
mean-field theory for the Hubbard-Peierls model proposed by
Zaanen and Littlwood~\cite{Zaanen2} suggests a picture in which
holes bind to domain walls embedded in an AF background.

Several years ago Zaanen and coworkers had introduced~\cite{Zaanen} an
appealing model for stripes acting as anti-phase boundaries intervening
AF spin domains, which captures well the essential features of the
interplay between spin and charge degrees of freedom. This
effective
model, deduced on basis of the Ogata-Shiba principle~\cite{Ogata} is
described in terms of fluctuating quantum strings. Zaanen's model can be
viewed as a variant of the bond-alternated spin model with bond
randomness as mentioned below and is interesting in its own rights as a
new random quantum spin system. Our main purpose in the present work is
to investigate the correlation between the spin and charge orderings and
the tendency of the spontaneous ordering of stripes within the effective
model. In particular, we carry out a
high precision numerical investigation into the charge and spin
orderings by direct calculation of the order parameters at
zero temperature in the thermodynamic limit by means of the quantum
Monte Carlo (QMC) simulation. 
We find that a hidden spin long-range order (LRO)
is governed by a competition between the two trends caused by increasing
hole mobility: the enhancement of the 2D spin-spin correlation and the
reformation of a strong stripe order.

\begin{figure}[t]    
 \centerline{\resizebox{0.45\textwidth}{!}{\includegraphics{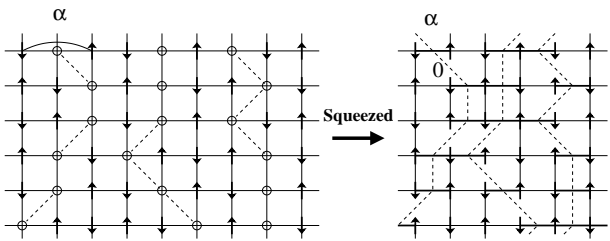}}} 
 \caption{Transformation from the original lattice to the squeezed
 lattice. The quantum strings in the original lattice are described by
 the dashed lines connecting links denoted by bold lines in the
 squeezed lattice.}
 \label{squeezed}
 \vspace{-0.5em}
\end{figure}

The motive of the aforementioned model is to describe the quantum
mechanical motion of hard core particles forming extended strings, which
fluctuate in an AF background. The strings have an overall orientation
which is aligned with the $y$-direction. Each
quantum string consists of $L_y$ hard-core particles referred to hereafter as
`holes'. Each hole moves in the $x$-direction and an effective exchange
interaction exists between the two spins neighboring a hole. By
regarding the hole as residing on the link between two spins neighboring
the hole, we can reduce our system to a spin-only model defined on a
squeezed lattice as depicted in Fig. \ref{squeezed}. 
The model thus obtained is described by the Hamiltonian~\cite{Zaanen}
\begin{eqnarray}
H & = & \sum_{x,y} [ -t P ( a^{\dagger}_{(x+1,y)} a_{(x,y)} + h.c. ) P \nonumber \\
  & + & J(1-(1-\alpha) n_{(x,y)}) {\bf S}_{(x,y)} \cdot {\bf S}_{(x+1,y)} \\
  & + & J(1-n_{(x,y)}n_{(x-1,y+1)}-n_{(x-1,y)}n_{(x,y+1)}) \nonumber \\
  & & \hspace*{0.1cm}{\bf S}_{(x,y)} \cdot {\bf S}_{(x,y+1)} ] \ , \nonumber
\end{eqnarray}
where the summation $\sum_{x,y}$ runs over all the lattice sites on the
squeezed lattice and ${\bf S}_{(x,y)}$ is the $S=1/2$ spin operator at
site $(x,y)$.
The first term corresponds to the kinetic energy of holes, where
$a^{\dagger}_{(x,y)}$ is the creation operator of the hole sitting on the
link between the sites ($x,y$) and ($x+1,y$) in the squeezed lattice,
and $P$ is the projection operator ensuring that strings are not
broken up and are separated by at least one spin site in the original
lattice. The second term describes the $S=1/2$ exchange interactions in
the $x$-direction. We fix $J$ to unity hereafter. The value of $\alpha$
corresponds to the effective
interaction between spins neighboring a hole and
$n_{(x,y)}$ is the hole number operator at site ($x, y$). The third term makes the
exchange interaction zero in a link parallel to $\hat{y}$ in the
squeezed lattice for bonds connecting to the holes in the original lattice. 

In the perfect stripe system with $t=0$, in which case the quantum
strings are perfectly straight and line up at regular intervals, 
the system as described on the squeezed lattice reduces to a spin ladder
model with an inter-ladder interaction of strength
$\alpha$~\cite{Tworzydlo}. When $L_x/N_{\rm s}$ is an odd
integer, where $L_x$ is the size of the lattice in the $x$
direction and
$N_{\rm s}$ is the number of strings, this system exhibits an
even-leg-ladder-like behavior:
there is a quantum phase transition between the disordered phase with a
finite spin gap for small $\alpha$ and the AF LRO phase for large 
$\alpha$~\cite{Matsumoto}. The transition point is located at e.g.
$\alpha_{\rm c}=0.3138(1)$, 0.0787(2), and 0.0153(1) 
for the concentrations $x=N_{\rm s}/L_x=1/3$, 1/5, and 1/7,
respectively. These results are obtained by the finite-size scaling
(FSS) analysis of our QMC data of the correlation length with the
exponent $\nu=0.71$ fixed to the value of the three-dimensional
classical Heisenberg universality class~\cite{Chen}. This disordered phase 
persists in the presence of the hopping
interaction up to a certain value of $t_{\rm c}$ as mentioned below. On
the other hand, when the value of $L_x/N_{\rm s}$ is even, the system
behaves like an odd-leg spin ladder: the system with $\alpha=0$ is in
a gapless phase and AF LRO is induced by an infinitesimal value of
$\alpha$. 

The strings in the present model, subject to quantum fluctuation viz. the 
$t$-term, are at the same time in contact with the AF background, and thus
induces an imaginary-time dependence of the spin exchange coupling on
the squeezed lattice. To deal with this aspect, the spin configuration
is therefore updated by the {\it discontinuous} 
imaginary-time loop algorithm~\cite{QMC} with fixed hole configuration. 
Meanwhile, the hole configuration is described by the Suzuki-Trotter
decomposition via the prescription due to Eskes {\it et
al}.~\cite{Eskes} and updated by the Metropolis algorithm with a fixed
spin configuration. The QMC simulation with the loop algorithm combined
with the conventional world-line algorithm is carried out on $L_x
\times L_y$ ($L_x = L_y \equiv L \le 36$) square lattices with the periodic boundary
condition. The value of $\beta/N_{\tau}$ is fixed to be 1/4,
where $\beta$ is the inverse of temperature $T$ and $N_{\tau}$ is the
Trotter number. The error bars are estimated based on about $10^2$
samplings. For each sample, $10^4$ Monte Carlo steps (MCS) are spent for
measurement after $10^4$ MCS for thermalization. 

\begin{figure}[t]    
 \centerline{\resizebox{0.28\textwidth}{!}{\includegraphics{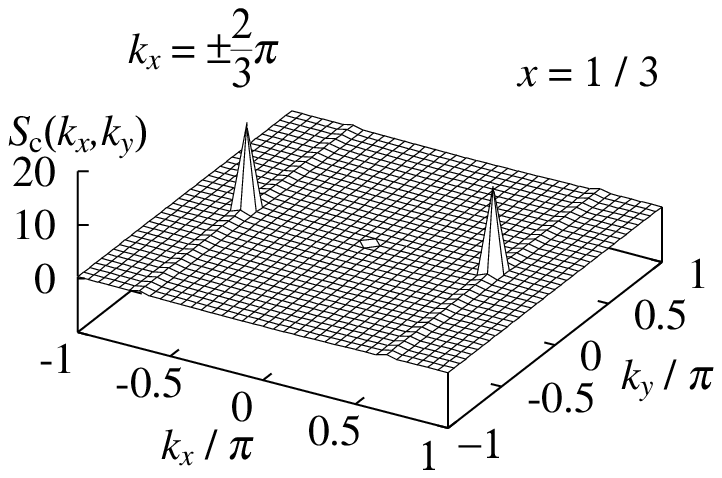}} \hspace{-8mm}
 \resizebox{0.28\textwidth}{!}{\includegraphics{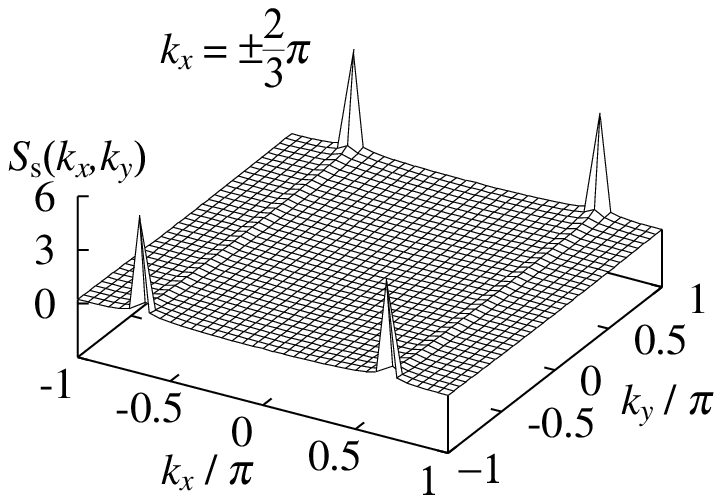}}}
 \vspace*{-1em}
 (a) \hspace{4cm}(b)
 \vspace{-0.5em}
 \caption{Plots of (a) the charge-charge structure factor and (b) the
 spin structure factor in the original lattice for $x=1/3$ at $T=0.01$
 in the charge-ordered and hidden AF long-range ordered phase with
 $\alpha=0.1$ and $t=1$. Note that the peak at (0,0) of the
 charge-charge structure factor is subtracted.}
 \label{sf1/3}
\vspace{-0.5em}
 \centerline{\resizebox{0.28\textwidth}{!}{\includegraphics{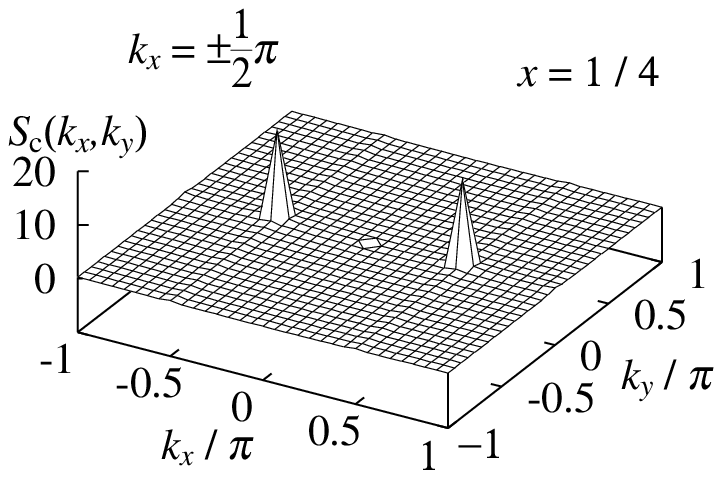}}  \hspace{-8mm}
 \resizebox{0.28\textwidth}{!}{\includegraphics{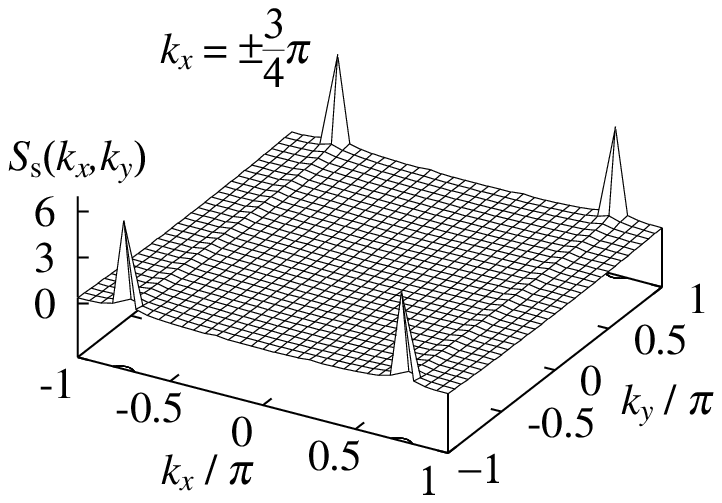}}} 
 \vspace*{-1em}
 (a) \hspace{4cm}(b)
 \vspace{-0.5em}
 \caption{Plots of (a) the charge-charge structure factor and (b) the
 spin structure factor in the original lattice for $x=1/4$ at $T=0.01$
 in the $\alpha=0.1$ and $t=1$ system.} 
 \label{sf1/4}
 \vspace{-0.5em}
\end{figure}

The ground-state phase diagram parameterized by $\alpha$ and $t$ for
$x=1/3$ was previously presented by Zaanen {\it et al}.~\cite{Zaanen}.
Our result is qualitatively consistent with theirs: the system is in
the charge-ordered and spin-disordered phase for small $t$ and small
$\alpha$ and the phase transition to the charge-ordered and hidden AF
LRO phase occurs as the value of $t$ or $\alpha$ is
increased. Hereafter, the AF LRO in the squeezed lattice will also be
referred to as the hidden AF LRO in terms of the original lattice. 
The spin-disordered state is effectively the same as the state realized
in the spin-1/2 even-leg-ladder system with the inter-ladder interaction
$\alpha$ as mentioned above. When $\alpha=0.1$ and $x=1/3$, our FSS analysis
of the Binder parameter~\cite{Binder} shows that the phase transition occurs at
$t_{\rm c}=0.21(1)$. The charge-ordered and hidden AF LRO
phase for $t>t_{\rm c}$ is characterized by peaks of the
charge-charge structure factor $S_{\rm c}(k_x, k_y) \equiv \frac{1}{L^2}
\langle [\sum_{\bf r}e^{i {\bf k} \cdot {\bf r}} n_{\bf r}]^2 \rangle$ at
($k_x,k_y$)=($\pm\frac{2}{3}\pi, 0$), where $n_{\bf r}$ is the 
hole number operator at site ${\bf r}$, and those of the spin structure
factor $S_{\rm s}(k_x, k_y) \equiv \frac{1}{L(L-N_{\rm s})}
\langle [\sum_{\bf r}e^{i {\bf k} \cdot {\bf r}} S_{\bf r}^z]^2 \rangle$
at ($k_x, k_y$)=($\pm\frac{2}{3}\pi,\pm\pi$) as shown in
Figs. \ref{sf1/3} (a) and (b). The results show that the hidden AF
correlations with anti-phase boundaries are 
present, consistent with the stripe structure, namely, the
peaks of $S_{\rm c}(k_x, k_y)$ and $S_{\rm s}(k_x, k_y)$
are at ($k_x, k_y$)=($\pm2\pi x,0$) and ($\pm(1-x)\pi,\pm\pi$),
respectively. Note that there are no additional peaks corresponding to
other modes, except for the one at
($k_x,k_y$)=(0,0) for $S_{\rm c}(k_x,k_y)$ in the thermodynamic limit.
While the $x=\frac{1}{3}$ system exhibits a even-leg-ladder-like
behavior, the system at $x=\frac{1}{4}$ is odd-leg-ladder-like. 
This is evident from the peaks that are observed at the positions 
($k_x,k_y$)=($\pm\frac{1}{2}\pi, 0$) and ($\pm\frac{3}{4}\pi,\pm\pi$)
as shown in Figs. \ref{sf1/4} (a) and (b), respectively.
It shows that at least in the parameter range we calculated, the
stripe configuration consisting of only three-leg-ladders is
realized rather than 
an alternating series of two- and four-leg-ladders, which was 
suggested in Ref. \cite{Zaanen} where interactions from holes were neglected.
The peaks of $S_{\rm c}(k_x, k_y)$ at $k_y=0$ shows that each string
selects a overall direction in space, though it can locally
fluctuate. This is associated with the ^^ ^^ directedness" observed in a
quantum string model~\cite{Eskes}. On the other hand, the peaks at
$k_x=\pm2\pi x$ corresponding to strings lined up at regular intervals might 
have their primary origin in the effective repulsion coming from the 
constraint that the stripes be separated by at least one spin site.

\begin{figure}[t] 
 \centerline{\resizebox{0.4\textwidth}{!}{\includegraphics{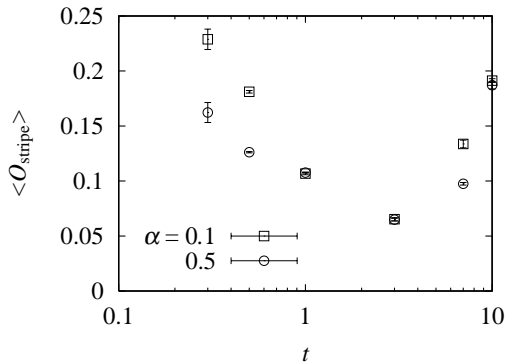}}}
 \vspace*{-1em}
\caption{The $t$-dependence of the stripe LRO parameter at $T=0$ for
 $x=\frac{1}{3}$. The square and circle denote the QMC results of the stripe
 LRO parameter for $\alpha=0.1$ and 0.5, respectively. }
 \label{stripe}
 \vspace{-0.5em}
\end{figure}

We make direct estimations of the $t$-dependent stripe LRO parameter
defined by 
\begin{equation}
   \langle O_{\rm stripe} \rangle \equiv \lim_{L \to \infty}
   \lim_{T \to 0} \sqrt{\frac{S_{\rm c}(2\pi x,0)}{L^2}} \ .
\end{equation}
The value of $S_{\rm c}(2\pi x,0)$ converges to its zero-temperature
value at $T$ lower than 
an energy scale related to the finiteness of the system size.
Thus, $S_{\rm c}(2\pi x,0)$ at low temperatures where its 
$T$-dependence becomes indiscernible within the error bars is taken as the
values in the $T \to 0$ limit. Furthermore, the value in the limit of $L
\to \infty$ is obtained by fitting the QMC data to $\sqrt{S_{\rm c}(2\pi
x,0)/L^2} \simeq \langle O_{\rm stripe} \rangle + a/L$.
The results of the extrapolation are shown in Fig. \ref{stripe}. When
the system is in the perfect stripe phase with $t=0$, the value of
$\langle O_{\rm stripe} \rangle$ equals $x$, because the strings are
perfectly straight and align at regular intervals, from which follows
that $S_{\rm c}(2\pi x,0)=N_{\rm s}^2$. For $t<3$ the value of $\langle
O_{\rm stripe} \rangle$ reduces as $t$ is increased. This is consistent
with the picture that hole fluctuation competes with the stripe order. 
For $t>3$, however, the strong stripe order is reformed and the ground
state approaches the perfect stripe. This reformation of the strong
stripe will be discussed later.

\begin{figure}[t] 
 \centerline{\resizebox{0.4\textwidth}{!}{\includegraphics{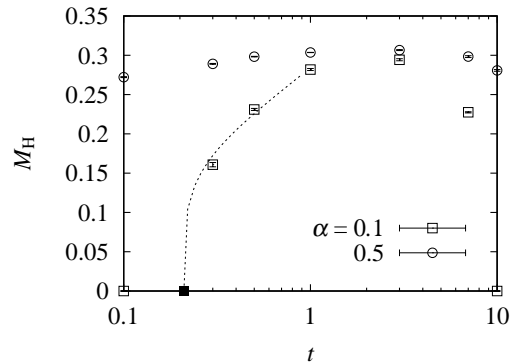}}}
 \vspace*{-1em}
\caption{The $t$-dependence of the hidden AF LRO
 parameter at $T=0$ for $x=\frac{1}{3}$. The open square and circle
 denote the QMC results of the hidden AF LRO parameter for $\alpha=0.1$ and
 0.5, respectively. The filled square denotes the critical point 
 $t_{\rm c}\simeq 0.21$ for $\alpha=0.1$. The dashed line is obtained by
 least-squares fitting with the fitting function 
$M_{\rm H}=c(x-0.21)^b$ in the range of $0.3 \le t \le 1$. The values of
 $c$ and $b$ are estimated to be 0.30(1) and 0.23(4), respectively.}
 \label{hidden}
 \vspace{-0.5em}
\end{figure}

We also make direct estimations of the $t$-dependent
hidden AF LRO parameter defined by 
\begin{equation}
   M_{\rm H} \equiv \lim_{L \to \infty} \lim_{T \to 0}
   \sqrt{\frac{3S(\pi,\pi)}{L(L-N_{\rm s})}} \ , 
\end{equation}
where $S(\pi,\pi)$ is the size-dependent staggered structure factor in
the squeezed lattice. The extrapolation is performed by 
a procedure similar to that 
for 
$\langle O_{\rm stripe} \rangle$. 
A finite $M_{\rm H}$ is induced at $t > t_{\rm c}=0.21(1)$ for
$\alpha=0.1$, with the value increasing with $t$ in the region $t_{\rm
c} < t \le 3$ as shown in Fig. \ref{hidden}. This result suggests that
the hole motion enhances the effective spin-spin interactions, which
in turn obscures the ladder-like structure, and
yields instead the 2D magnetic LRO.
To paraphrase, the magnetization which had been suppressed 
in the underlying bond-alternating spin system due to quantum fluctuations 
is recovered by the randomness of the spin interaction brought on by turning 
on a finite $t$. Randomness-induced-LRO is expected to  
occur also in 2D quantum spin systems with  bond randomness~\cite{Fisher}. The 
present case differs in that the randomness is present also 
in the imaginary-time direction, i.e., one may view it as a 
^^ ^^ 2+1D random quantum spin system".
A further increase in the hole motions, however, eventually leads to the
reduction of the spin LRO due to the reformation of the strong stripe order,
leading to the spin system with alternating structure which was seen for
$0<t<t_{\rm c}$. The properties mentioned above are also seen for
$\alpha=0.5$, where the system with $t \to 0$ does not belong to the
spin-disordered phase. 

The values of $\langle O_{\rm stripe} \rangle$ for $\alpha=0.1$ are
larger than those for $\alpha=0.5$ as shown in Fig. \ref{stripe}. This
result is in conflict with the following naive energetics: a kink,
arising when one hole in the straight string moves to a neighboring
site, yields an energy loss 
proportional to a nearest spin-spin correlation.
In the hidden AF LRO phase, a naive mean-field-like argument gives the
correlation proportional to $M_{\rm H}^2$.
This implies then that in order to gain energy the value of the
stripe LRO parameter for $\alpha=0.5$ should become larger than that for
$\alpha=0.1$ because of the $\alpha$-dependence of $M_{\rm H}$. Our results indicate that the speculation is too naive to
explain quantitatively the magnitude of the order parameters.

\begin{figure}[t]
 \centerline{\resizebox{0.4\textwidth}{!}{\includegraphics{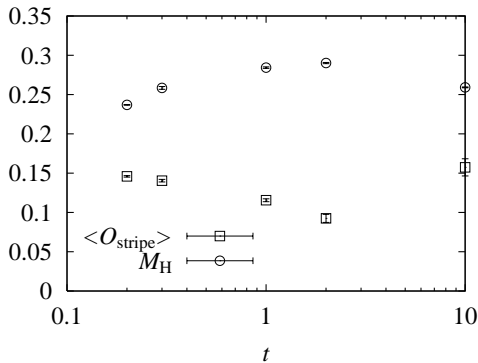}}}
 \vspace*{-1em}
\caption{The $t$-dependence of the LRO parameters at $T=0$ for
 $x=\frac{1}{4}$. The square and circle denote the values of the stripe
 LRO and the hidden AF LRO parameters for $\alpha=0.1$.}
 \label{x=1/4}
 \vspace{-0.5em}
\end{figure}

The $t$-dependences of $\langle O_{\rm stripe} \rangle$ and $M_{\rm H}$
for $\alpha=0.1$ and $x=\frac{1}{4}$ are also shown in Fig. \ref{x=1/4},
where the system in the limit of $t \to 0$ becomes odd-leg-ladder-like.
Though a similar upturn behavior of the order parameters to that for
$x=1/3$ is observed, the $t$-dependence is indiscernible due to
the weak effective repulsion between quantum strings. 
We thus expect in general, that it becomes harder in practice 
to see the correlation between spin and charge orderings as one goes to
lower hole densities, in particular for the odd-leg-ladder like system.

To conclude, we have demonstrated that the magnitude of the hidden AF LRO is
governed by the competition between the enhancement of the 2D spin-spin
correlation and the reformation of the strong stripe order.
The question as to why the stripe tends to reorder at large $t$ remains
to be resolved. Although we have not yet found a rigorous description of the 
reformation of the strong stripe order, 
it may be due to the effective Coulomb repulsion coming from
the constraint that the stripes are separated by at least one spin
site. The present model with the constraint is
not realistic at large $t$.
There are two 
further issues along the line of the present study left for
future work. One is the effect of adding a next-nearest neighbor
interaction on the squeezed lattice. In that case, the diagonal stripe,
observed in Cu-oxides and Ni-oxides, would be expected to emerge 
in a certain parameter region. The other is to incorporate the
dislocation of the strings. In the present work, we have restricted the
string mobility so as to ensure that the strings are not broken up. In 
order to access a superconducting regime, however, it would be crucial
to study the situation where the stripes start to get
destroyed~\cite{Zaanen}. The numerical simulation for this situation is
a major challenge and new ideas are in need.

We acknowledge fruitful discussions with T. Hikihara and M. Miyazaki.
The numerical calculations in the present work have been performed
on the ITBL super-computer system. The ``PARAPACK ver. 2'' developed by
S. Todo is used for parallel computation.
This study is partially supported by the Ministry of
Education, Culture, Sports, Science and Technology, Japan, under the
Priority Grant No. 14038240.


\vspace*{-1.5em}
\begin{thebibliography}{99}
\vspace*{-1.5em}
\bibitem{Tranquada1} J. M. Tranquada {\it et al}., Nature {\bf 375}, 561
	(1995); J. M. Tranquada {\it et al}., Phys. Rev. B {\bf 54},
	7489 (1996).
\bibitem{Noda} T. Noda, H. Eisaki, and S. Uchida, Science {\bf 286}, 265
	(1999). 
\bibitem{Zhou} X. J. Zhou {\it et al}., Science {\bf 286}, 268 (1999).
\bibitem{Cheong} S-W. Cheong {\it et al}., Phys. Rev. Lett. {\bf 67},
	1791 (1991).
\bibitem{Ino} A. Ino {\it et al}., J. Phys. Soc. Jpn. {\bf 68}, 1496 (1999).
\bibitem{Machida} K. Machida, Physica C {\bf 158}, 192 (1989); M. Kato {\it et al}., J. Phys. Soc. Jpn. {\bf 59}, 1047 (1990).
\bibitem{Poilblanc} D. Poilblanc and T. M. Rice, Phys. Rev. B {\bf 39}, 9749 (1989).
\bibitem{Zaanen3} J. Zaanen and O. Gunnarsson, Phys. Rev. B {\bf 40}, 7391 (1989).
\bibitem{Schulz} H. J. Schulz, Phys. Rev. Lett. {\bf 64}, 1445 (1990).
\bibitem{Giamarchi} T. Giamarchi and C. Lhuillier, Phys. Rev. B {\bf
	42}, 10641 (1990). 
\bibitem{Miyazaki} M. Miyazaki, K. Yamaji, and T. Yanagisawa,
	J. Phys. Chem. Solids, {\bf 63}, 1403 (2002).
\bibitem{White} S. R. White and D. J. Scalapino, Phys. Rev. Lett. {\bf
	80}, 1272 (1998).
\bibitem{Sakai} T. Sakai, Phys. Rev. B {\bf 63}, 140509(R) (2001).
\bibitem{CChen} C. H. Chen, S-W. Cheong, and A. S. Cooper,
	Phys. Rev. Lett. {\bf 71}, 2461 (1993).
\bibitem{Tranquada3} J. M. Tranquada {\it et al}., Phys. Rev. Lett. {\bf
	73}, 1003 (1994); J. M. Tranquada, D. J. Buttrey, and V. Sachan,
	Phys. Rev. B {\bf 54}, 12318 (1996).
\bibitem{Zaanen2} J. Zaanen and P. B. Littlewood, Phys. Rev. B {\bf 50},
	R7222 (1994).
\bibitem{Zaanen} J. Zaanen {\it et al}., Philosophical Magazine B {\bf
	81}, 1485 (2001).
\bibitem{Ogata} M. Ogata and H. Shiba, Phys. Rev. B {\bf 41}, 2326
	(1990). 
\bibitem{Tworzydlo} J. Tworzyd{\l}o {\it et al}., Phys. Rev. B {\bf 59},
	115 (1999).
\bibitem{Matsumoto} M. Matsumoto {\it et al}., Phys. Rev. B {\bf 65},
	014407 (2002). 
\bibitem{Chen} K. Chen, A. M. Ferrenberg, and D. P. Landau, Phys. Rev. B
        {\bf 48}, 3249 (1993).
\bibitem{QMC} H. G. Evertz, G. Lana, and M. Marcu, Phys. Rev. Lett. {\bf
        70}, 875 (1993); B. B. Beard and U.-J. Wiese,
        {\it ibid.} {\bf 77}, 5130 (1996); S. Todo and K. Kato,
        {\it ibid.} {\bf 87}, 047203 (2001).
\bibitem{Eskes} H. Eskes {\it et al}, Phys. Rev. B {\bf 58}, 6963 (1998).
\bibitem{Binder} K. Binder, Z. Phys. B {\bf 43}, 119 (1981).
\bibitem{Fisher} D. S. Fisher, Phys. Rev. B {\bf 50}, 3799 (1994).
\end{thebibliography}
\end{document}